\documentclass{jetpl}
\twocolumn
\lat

\title{Formation of exotic states in the  $s-d$ exchange and $t-J$  models}
\rtitle{}

\sodtitle{}

\author{V. Yu. Irkhin
\/\thanks{e-mail: valentin.irkhin@imp.uran.ru} and Yu. N. Skryabin}

\rauthor{V. Yu. Irkhin, Yu. N. Skryabin}

\sodauthor{Irkhin, Skryabin}

\address{M. N. Mikheev Institute of Metal Physics, 620108 Ekaterinburg, Russia}

\abstract{Different scenarios of the implementation of the two-band model in strongly correlated electrons systems, including frustrated magnets, high-temperature superconductors, and Kondo lattices, are considered. The interaction of current carriers with magnetic moments in the representations of pseudofermions or Schwinger bosons describing the spinon excitations is studied on the basis of the derived Hamiltonians of the $s-d$ exchange and $t-J$ models within the formalism of many-electron Hubbard operators.
}


\begin{document}

\maketitle

\textbf{1. Introduction}

Unusual excitations and exotic states in strongly
correlated solid states and other condensed media,
e.g., different types of spin liquids and states with
topological and quantum orders, are currently under
active investigation [1]. They are usually described
within many-electron models. The simplest of them is
the one-band t-J model, in which strong single-site
correlations and the exchange interaction between
localized spins are taken into account. Though it
makes it possible to consider a series of exotic phases
and is successively applied to physics of cuprates (basic
systems for high-temperature superconductors
(HTSCs)) [2, 3], the corresponding approximations
are difficult to control owing to the absence of a small
parameter: the simplicity of a model often does not
provide the convenience of the theoretical study.

On the other hand, in physics of magnetic semiconductors,
heavy fermion systems, and Kondo lattices,
as a rule, one uses the two-band s-d(f) exchange
model, in which subsystems of current carriers and
local magnetic moments are separated initially; in
addition, the model is convenient because it has a
semiclassical small parameter. The idea of such separation
acquires a new meaning in current field-theoretical
approaches, where the formation of exotic
phases and particles with unusual statistics is considered.
It is assumed that such states are implemented in
insulating and conducting f- and d-systems (here, in
addition to cuprates, frustrated band magnets of the
YMn$_2$ type can be mentioned [4]).

In this work, we show that the usage of the two-band
(in particular, s-d exchange) model as the more
general one makes it possible to clarify a series of
moments in the physical understanding of the state of
spin liquid. In addition, we demonstrate the description
of exotic excitations in terms of many-electron
Hubbard operators.
$\newline$  $\newline$  $\newline$   
$\newline$
\textbf{2. Model Hamiltonians}

The t-J model, which is the Hubbard model with
the infinite single-site repulsion $U\rightarrow \infty $ and allowance
for the Heisenberg exchange, is widely used at the
theoretical consideration of strongly correlated compounds
(e.g., copper-oxygen HTSC). Its Hamiltonian
in the many-electron representation of Hubbard
operators
$X(\Gamma ,\Gamma ^{\prime })=|\Gamma \rangle \langle \Gamma
^{\prime }|$ takes the form

\begin{equation}
H=\sum_{ij\sigma }t_{ij}X_i(0\sigma )X_j(\sigma 0)
+ H_d, \, H_d = \sum_{ij}J_{ij}\mathbf{S}_i\mathbf{S}_j.
 \label{eq:I.7}
\end{equation}

The model (1) describes the interaction of current
carriers with local moments. To demonstrate explicitly
the separation of these degrees of freedom, we note its
equivalence to the s-d exchange model with the
exchange parameter $I \rightarrow -\infty $. In fact,
after passing to
the many-electron representation, the Hamiltonian of
the latter is written as [5, 6]
\begin{equation}
H=\sum_{ij\sigma
}t_{ij}g_{i\sigma  }^{\dagger
}g_{j\sigma}+H_d,
 \label{eq:I.5}
\end{equation}
\begin{equation}
g_{i\sigma }^{\dagger }=\sum_M C_{SM;1/2, \sigma}^{S-1/2,M+\sigma}
X_i(S-1/2, M+ \sigma;SM),
\end{equation}
where
\begin{equation}
 C_{SM;1/2, \sigma}^{S-1/2,M+\sigma}=(S-2\sigma M)^{1/2}/(2S+1)^{1/2}
\end{equation}
are the Clebsch-Gordan coefficients for the coupling
of spins $S$  and 1/2. It is easy to see that the Hamiltonian
(2) at coincides with (1) with the trivial
renormalization: $t_{\mathbf{k}}$ is replaced in (2) by $2t_{\mathbf{k}}$ (the factor 2
appears owing to the equivalence of transitions of electrons
with both opposite spins in the Hubbard model).

The standard representation of auxiliary (slave)
bosons introduced by Anderson [2] has the form
\begin{equation}
X_i(\sigma ,0)
=f_{i\sigma }^{\dagger}e_i
 \label{eq:6.131}
\end{equation}
Here, $e_i$ are annihilation operators for charged spinless
bosons (holons) and $f_{i\sigma }^{\dagger }$ are creation operators for
neutral fermions (spinîns). It is also possible to use
the representation of auxiliary fermions
\begin{equation}
X_i(\sigma ,0)
= b_{i\sigma }^{\dagger}e_i,
\label{eq:6.11}
\end{equation}
where now $e_i$  and are Fermi and $b_{i\sigma }^{\dagger }$ Bose (Schwinger)
operators, respectively. However, we will use directly
the X--operators, introducing spinons only for the
localized subsystem.

In turn, the Hamiltonian (2) can be expressed in
terms of Fermi operators and operators of localized
spins. Using the representation of Hubbard operators
in terms of many-electron operators of the creation of
electron configurations $ A_\Gamma ^{\dagger}$ [5, 6],

\begin{equation}
X(\Gamma ,\Gamma ^{\prime })=A_\Gamma ^{\dagger}\prod_\sigma
(1-\ n_\sigma) A_{\Gamma ^{\prime }}, \quad n_\sigma= c_\sigma ^{\dagger } c_\sigma
 \label{eq:A.21}
 \end{equation}
and the relation
\begin{equation}
A_{S-1/2,\mu}^{\dagger }=\sum_{M\sigma}C_{SM,1/2\sigma }^{S-1/2,\mu }c_\sigma ^{\dagger }A_{SM}^{\dagger }, \quad .
 \label{eq:A.12}
\end{equation}
it is possible to separate the operators of conduction
electrons on the space of singly occupied states from
X--operators. Identically transforming the product of
Clebsch-Gordan coefficients, we find
\begin{equation}
g_{i\sigma }^{\dagger }=\sum_{\sigma ^{\prime }}c_{i\sigma
^{\prime }}^{\dagger }(1-n_{i,-\sigma ^{\prime }})
\frac {S \delta _{\sigma \sigma ^{\prime }}-(\mathbf{S}_i\mbox{\boldmath$\sigma $}_{\sigma ^{\prime }\sigma })} {2S+1}.
 \label{eq:I.8}
\end{equation}
Further,  using the properties of Pauli matrices
gives
\begin{multline}
H=\frac{1}{(2S+1)^{2}}\sum_{ij\sigma \sigma ^{\prime }}t_{ij}\{(S^{2}+%
\mathbf{S}_{i}\mathbf{S}_{j})\delta _{\sigma \sigma ^{\prime }}-S(\mathbf{S}%
_{i}+\mathbf{S}_{j})\mbox{\boldmath$\sigma $}_{\sigma \sigma ^{\prime }} \\
+i\mbox{\boldmath$\sigma $}_{\sigma \sigma ^{\prime }}[\mathbf{S}_{i}\times
\mathbf{S}_{j}]\}c_{i\sigma }^{\dagger }(1-n_{i,-\sigma })(1-n_{j,-\sigma
^{\prime }})c_{j\sigma ^{\prime }}+H_{d}.  \label{eq:I.10}
\end{multline}
Such representation of the Hamiltonian (of course,
also valid in the t-J model) was obtained for the first
time in [5, 6]. Later, it was used in [7] as applied to the
phase diagram of HTSC cuprates as ``new formulation
of the t-J model''. a somewhat different interpretation
being given: the arising electron states were called
dopons.

The transition to the s-d exchange model with
expand the physical space and at the same time (unlike
the consideration of the t-J model [7]) remove the
nonphysical state owing to the condition $I \rightarrow -\infty$.
However, the generalization to the case of the s-d
exchange model with finite $I$ is also possible:
\begin{equation}
H=\sum_{{\bf k}\sigma }t_{{\bf k}}c_{{\bf k}\sigma }^{\dagger }c_{{\bf k}%
\sigma }-I\sum_{i\alpha \beta }{\bf S}_i\mbox {\boldmath $\sigma $}_{\alpha
\beta }c_{i\alpha }^{\dagger }c_{i\beta }+H_d,
\label{finite}
\end{equation}
which, in particular, describes Kondo lattices.

Terms linear in spin operators in Eq. (10) provide
the possibility of the effective hybridization of electrons
(``dopons'' according to Ref.7)
with spinons, as well as in Kondo lattices. Terms containing
vector products, though they disappear for
simple spin configurations, describe the anisotropic
scattering of electrons. Thereby, the Hamiltonian (10)
can be useful in the consideration of states with anomalous
``chiral'' order parameters of kinetic phenomena
in narrow bands, e.g., the anomalous Hall effect (see,
e.g., [8]).
     
     $\newline$
\textbf{3. Case of a single current carrier}

In model (11), the expression for the energy of the
electron in the second-order perturbation theory in
the case of the empty conduction band has the form
\begin{equation}
\Sigma^{(2)} _{\mathbf{k}}(E)=\Phi _{\mathbf{k}}(E)=I^{2}\sum_{\mathbf{q}}\int \frac{K_{\mathbf{q}}(\omega
)d\omega}{E-t_{\mathbf{k+q}}+\omega }  \label{S}
\end{equation}%
where  $K_{\mathbf{q}}(\omega)$ is the spin spectral function. To find it,
we consider different spinon representations for localized
spins.

In the self-consistent spin-wave theory, the representation
of Schwinger bosons is used [9]:
\begin{equation}
\mathbf{S}_i=\frac 12\sum_{\sigma \sigma ^{\prime }}b_{i\sigma
}^{\dagger }\mbox{\boldmath$\sigma $}_{\sigma \sigma ^{\prime
}}b_{i\sigma ^{\prime }}
 \label{eq:O.1}
\end{equation}
or that of Dyson-Maleev [10]. This makes it possible
to describe the quantum disordered state (spin liquid)
and the magnetically ordered phase with the wave vector
$\mathbf{Q}$, the latter being considered as the boson condensate
(at low temperatures, the state close to the
condensate arises in the regions with the exponentially
large correlation length, which is described analogously;
i.e., the ``effective'' magnetization of the sublattice
 $\bar{S}_{\text{eff}}(T)$
does not practically change [11]). After the separation
of condensate contributions (extra factors of 3/2
appear in them when using the Schwinger bosons,
which we omit), we find
\begin{multline}
K_{\mathbf{q}}(\omega )=\overline{S}_{\text{eff}}^{2}\delta (\omega )\delta
_{\mathbf{qQ}}+\overline{S}_{\text{eff}}^{{}}(u_{\mathbf{q}}-v_{\mathbf{q}%
})^{2}\delta (\omega
+\omega _{\mathbf{q}})\\
+\sum_{\mathbf{p}}(u_{%
\mathbf{p-q}}v_{\mathbf{p}}-v_{\mathbf{p-q}}u_{\mathbf{p}})^{2}\delta
(\omega +\omega _{\mathbf{p-q}}+\omega _{\mathbf{p}})  \label{K}
\end{multline}%
The first term in (14) leads to the formation of the
antiferromagnetic gap and will be ignored (this is justified
near the band bottom). The second term
describes the interaction with spin waves (it is absent
in the quantum disordered state, since a gap arises in
the spectrum of spinons), and the third term describes
the interaction with the individual spinon excitations.

In the state of resonating valence bonds, the Hamiltonian
of the -subsystem is written in the representation
of pseudofermions analogous to (13) as in [3,
12]:
\begin{equation}
H_{d}=\sum_{\mathbf{k}\sigma }(B_{\mathbf{k}}f_{\mathbf{k}\sigma }^{\dag }f_{%
\mathbf{k}\sigma }^{{}}+\Delta _{\mathbf{k}}f_{\mathbf{k}\sigma }^{{}}f_{-%
\mathbf{k-}\sigma }^{{}}+h.c.\},  \label{hferm}
\end{equation}%
with $B_{\mathbf{k}} \sim J_{\mathbf{k}}$. The corresponding spectral density is obtained in
the approximation of noninteracting Fermi spinons
and is analogous to the last term in Eq. (14):
\begin{equation}
K_{\mathbf{q}}(\omega )=\sum_{\mathbf{k}}(u_{\mathbf{k-q}}v_{\mathbf{%
k}}-v_{\mathbf{k-q}}u_{\mathbf{k}})^{2}\delta (\omega +E_{\mathbf{k-q}}+E_{%
\mathbf{k}})  \label{KF}
\end{equation}%
where the spectrum
$E_{\mathbf{k}}=(B_{\mathbf{k}}^2+\Delta_{\mathbf{k}}^2)^{1/2}$ and the coefficients
of the Bogoliubov transformation (now Fermi
type) are obtained during the diagonalization of (15).

In the antiferromagnetic and quantum disordered
states and in the resonating valence bond state, the
combination of coefficients of the $u-v$ transformation
in $K_{\mathbf{q}}(\omega )$ becomes zero at $q \rightarrow 0$.

We consider the structure of the electron spectrum
near the band bottom in the case of a single current
carrier. In the self-consistent approximation, replacing
energy denominators by exact Green's functions,
we obtain the integral equation
\begin{equation}
\Phi _{\mathbf{k}}(E)=I^2\sum_{\mathbf{q}}
 \int K_{\mathbf{q}}(\omega
)G_{\mathbf{k}+\mathbf{q}}(E+\omega )\,d \omega .
 \label{eq:6.114}
\end{equation}
To solve Eq. (17), it is possible to apply the ``dominating pole''
approximation  [11, 12]:
\begin{equation}
G_{\mathbf{k}}(E)=\frac{Z_{\mathbf{k}}}
{E-\tilde{E}_{\mathbf{k}}}+G_{\text{inc}}(\mathbf{k},E),
 \label{eq:6.116}
\end{equation}
where  $G_{\text{inc}}$ is the incoherent contribution to the
Green's function and
\begin{equation}
Z_{\mathbf{k}}=\left( 1-\frac \partial {\partial E}{\rm Re} \Sigma
_{\mathbf{k}}(E)\right) _{E=\tilde{E}_{\mathbf{k}}}^{-1}
 \label{eq:6.117}
\end{equation}
is the residue at the pole near the band bottom corresponding
to the spectrum of new quasiparticles $\tilde{E}_{\mathbf{k}}$.

Substituting Eq. (18) into Eq. (17) and integrating
over $q$, we obtain the estimate in the two-dimensional
case:
\begin{equation}
Z^{-1}-1\sim I^2/|Jt|.
 \label{eq:6.119}
\end{equation}
Thus, with  increasing $|I|$ , the spectral weight
transfers into the incoherent contribution and
undamped quasiparticles become heavy, so that, at
$I^2\gg J|t| $ , the effective mass is given by  $m^{*}/m=Z^{-1}\gg 1$. In
the three-dimensional case, the divergence is weaker,
and corrections to the residue contain only the logarithmic
factor: $Z^{-1}-1 \sim I^2S \ln|t/JS|$ [11].

In the case of narrow bands ($I \rightarrow -\infty $), we consider
the Green's function of many-electron operators
\begin{equation}
G_{\mathbf{k}\sigma }(E)=\langle \!\langle g_{\mathbf{k}\sigma }|g_{\mathbf{k%
}\sigma }^{\dagger }\rangle \!\rangle _{E}.\quad
\end{equation}
The result of calculations with allowance for spin fluctuations
[11] has the form
\begin{equation}
G_{\mathbf{k}\sigma }(E)=\Psi _{\mathbf{k}}(E)/[E-\Psi _{\mathbf{k}}(E)t_{%
\mathbf{k}}],  \label{eq:6.125}
\end{equation}
\begin{multline}
\Psi _{\mathbf{k}}(E)=\frac{S}{2S+1}+\\
\sum_{\mathbf{q}}\frac{t_{\mathbf{k}+%
\mathbf{q}}}{(2S+1)^{2}}\int K_{\mathbf{q}}(\omega )\Psi _{\mathbf{k}+%
\mathbf{q}}^{-1}(E)G_{\mathbf{k}+\mathbf{q}}(E+\omega )\,d\omega.
\label{eq:6.126}
\end{multline}%
As was noted in [12], the effect of the finite width of
the bare band (the first term on the right-hand side of
Eq. (23)) and details of the approximation are not
important for the quasiparticle pole, though they are
essential for the incoherent contribution. As a result,
we find for two- and three-dimensional cases
\begin{equation}
Z^{-1}-1\sim
\begin{cases}
|t/JS|,& D=2, \\
S^{-1}\ln{} |t/JS|, & D=3.
\end{cases}
 \label{eq:6.130}
\end{equation}

Our approach considers physical excitations rather
than auxiliary particles, i.e., holons (bosons or fermions)
as in [12] (where it is also noted that the calculation
of the boson Green's function is not quite physically
consistent). Since the results (18) and (20) are
determined by divergences at low momenta, they are
also obtained in the spin-wave picture [11] and in both
ways of the consideration of the quantum disordered
state in terms of bosons and fermions. We will see
below that a similar change of the statistics from the
Bose to Fermi one also occurs in Kondo lattices.

The further development of the theory takes into
account the interaction with the gauge field [3];
apparently, it is not too important in the case of the
spin liquid of the type $Z_2$, but is essential for U(1).

  $\newline$
\textbf{4.  Case of the finite band filling}

At the partial filling of the conduction band, we
come to the situation of the Kondo lattice. If the
Kondo temperature $T_K \sim \exp (1/2I\rho(E_F))$
(where $\rho (E)$
is the bare density of states) is much higher than
the d-d exchange  $J$, localized moments are screened
by conduction electrons and the ground state of the
heavy Fermi liquid appears. However, other exotic
phases can also appear in the general case. If the d-d
bonds are frustrated, the d-moments can form the spin
liquid, which does not violate the symmetry of the lattice
Hamiltonian [3]. The deconfinement state arises
at the boundary between the magnetic phase and the
Fermi liquid in the form of the algebraic spin liquid
with the non-Fermi-liquid behavior and the separation
of the charge and spin degrees of freedom [3]. The
formation of the exotic disordered state can be
achieved not only by the direct introduction of frustration
into the spin subsystems but also by doping.

First, we discuss the results of perturbation theory
[13-15]. The calculation of the magnetic susceptibility
with allowance for spin dynamics leads to the result
\begin{equation}
\chi =\frac{S(S+1)}{3T}-\frac{4I^2}{3T}\sum_{\mathbf{p}\mathbf{q}}\int
K_{\mathbf{p}-\mathbf{q}}(\omega )
\frac{n_{\mathbf{p}}(1-n_{\mathbf{q}})}
{(t_{\mathbf{q}}-t_{\mathbf{p}}-\omega )^2}\,d\omega,
 \label{eq:6.83}
\end{equation}
where the second term describes the screening of the
moment and $n_{\mathbf{k}}$ are Fermi distribution functions.
The Kondo contribution to the imaginary part of
the self-energy arising in the third-order perturbation
theory has the form
\begin{equation}
{\rm Im}\Sigma _{\mathbf{k}}^{(3)}(E)=2\pi I^3\rho (E)\int
\sum_{\mathbf{q}}K_{\mathbf{q}}(\omega
)\frac{n_{\mathbf{k}+\mathbf{q}}}
{E-t_{\mathbf{k}+\mathbf{q}}-\omega }\,d\omega
 \label{eq:6.31}
\end{equation}
The usage of the corresponding spectral function
makes it possible to take into account the cutoff of
Kondo divergences and to develop the renormalization
group theory of Kondo lattices analogously to [13,
14] not only in magnetic phases but also for states of
the spin liquid type with the non-Fermi-liquid
behavior.
For the description of the ground state (the strong
coupling regime), it is possible to use the mean field
theory in the representation of pseudofermions [16-
19]. The most important effect here is the formation of
the ``large'' Fermi surface, in which pseudofermions
hybridizing with conduction electrons are involved.
The condensation of Bose spinons (Schwinger
bosons) means the formation of magnetism, and the
condensation of the Higgs boson ($b_0 \sim \langle f_{i\sigma }^{\dagger}c_{i\sigma }\rangle$) in the Kondo phase means the formation of the Fermi liquid
with a large Fermi surface, though secondary magnetic
ordering is also possible in this phase [18, 19].

The formation of the spin liquid (an exotic FL$^*$ Fermi
liquid in the terminology of [18, 19]) is the intermediate
regime: the same as the magnetic ordering, it suppresses
the Kondo effect; therefore, the effective
value is not renormalized to infinity, as takes place in
the case of a single Kondo impurity. The s-electrons
are weakly coupled to the d-spin liquid and form the
``small'' Fermi surface, which covers the volume
determined only by the density of s-electrons.

In the case of narrow bands (the t-J model), doping
is responsible for a series of complex effects, in
particular, for the competition of ferro- and antiferromagnetism
leading to the formation of helical or nonuniform
magnetic structures; more exotic (including
topological) states, the formation of the pseudogap are
also possible [3].

The consideration of the spectrum of cuprates
reveals nodal-antinodal dichotomy [3]: the nature of
the spectrum differs in different regions of the Fermi
surface. The spectrum is gapless near the nodal points
$(\pm\pi/2,\pm\pi/2)$ (where the excitations are described as
Dirac fermions) and has a gap near the antinodal point  $(0,\pi)$.

In the mean field theory for the t-J model [7], the
electron spectral weight originates from two bands: the
low-energy spinon band and the high-energy electron
band. The spectral weight from the spinon band is a
sharp coherent peak. The broad spectral weight from
the electron band corresponds to the incoherent background.
The strong hybridization mixing between spinons
and electrons described by the Hamiltonian (10)
arises near the nodal point, and mixing is absent near
the antinodal point. The Fermi surface is large or
remains small, respectively. In essence, the described
picture is close to the hybridization two-band model of
Kondo lattices, where the separation of the localized
and itinerant states appears.

On the other hand, the representation of Schwinger
bosons was used in [20] to describe the state with a
small Fermi surface (spin liquid of the $Z_2$ type) in the
same model, though no separation of the coherent
band analogous to that considered in Section 3 was
performed (the corresponding perturbation theory in
the many-electron representation was earlier developed
in [11, 21]).

The approach of many-electron operators makes it
possible to simultaneously consider both cases and
take into account the change in the statistics of spinons
at the quantum phase transition, introducing different
spectral functions for collective excitations.

   $\newline$
\textbf{5. Concludions}

We see that the usage of the two-band model makes
it possible to provide the general physical description
of exotic states of strongly correlated systems.

In the case of the Hubbard model with the finite
Coulomb repulsion, we deal with the problem of the
formation of local magnetic moments. The presence
of the ``direct'' Heisenberg exchange favors their
appearance, so that the physical situation becomes
close to the s-d exchange model (formally, the Coulomb
term at one site can be written in the ``exchange''
form $-U\mathbf{s}_i \mathbf{s}_i$; besides that,  the Anderson's
superexchange $\sim (t^2_{ij}/U)\mathbf{s}_i \mathbf{s}_j$
occurs  at large $U$). The separation of contributions of the
coherent (quasiparticle) and incoherent (non-quasiparticle)
states is essential. The latter can be described
most simply in the ferromagnetic phase, where they,
however, arise only for the down spin projection,
describing bound states of the carrier with the spin up
and magnon [22, 23].

The representation of Schwinger bosons can be
introduced not only in the t-J model [20] but also in
the antiferromagnetic phase of the spin-fermion
model, which is the modification of the single-band
Hubbard model with finite interaction [24]. Therefore,
the exotic Fermi liquid FL$^*$ is implemented in the latter
model. The spin-Fermi model [25] used in the
interpolation description of collective magnetism (the
analogy of which with the s-d exchange model is
described in [26]) also makes it possible to separate
spin and electron degrees of freedom.

The ``Kondo'' peak on the Fermi level also appears
in approximation of the dynamic mean field theory
reducing the Hubbard model to the effective Anderson
model [27]. It should be noted that, unlike the Hubbard
model, where the Coulomb interaction leads to
the destruction of the Fermi liquid state (the formation
of Hubbard subbands with a small Fermi surface),
the effect of the s-d exchange is the opposite: a heavy
Fermi liquid and a large Fermi surface appear with the
increase in . The origin is that these interactions lead
to different pairing types of electrons and spinons,
diagonal and off-diagonal (hybridization).

The change in the statistics of spinons occurs at the
quantum phase transition between two phases of the
confinement: the magnetic phase and the phase with
the large Fermi surface. Such a problem requires further
studies. Here, supersymmetric representations
may turn out to be useful (see, e.g., [28,29]).

The research was supported by the Russian Science
Foundation (project no. 17-12-01207).

\end{document}